\documentclass[english,pra,amsmath]{revtex4}
\usepackage[T1]{fontenc}
\usepackage[latin9]{inputenc}
\usepackage{graphicx}

\usepackage{amsfonts}

\providecommand{\U}[1]{\protect\rule{.1in}{.1in}} 
\providecommand{\U}[1]{\protect\rule{.1in}{.1in}} \providecommand{\U}[1]{\protect\rule{.1in}{.1in}} \providecommand{\U}[1]{\protect\rule{.1in}{.1in}}
\providecommand{\U}[1]{\protect\rule{.1in}{.1in}} \providecommand{\U}[1]{\protect\rule{.1in}{.1in}} \providecommand{\U}[1]{\protect\rule{.1in}{.1in}}
\providecommand{\U}[1]{\protect\rule{.1in}{.1in}} \providecommand{\U}[1]{\protect\rule{.1in}{.1in}} \providecommand{\U}[1]{\protect\rule{.1in}{.1in}}
\providecommand{\U}[1]{\protect\rule{.1in}{.1in}}

\begin{document}

\author{H. Cable$^{a}$, F. Lalo\"{e}$^{b},$ and W. J. Mullin$^{c}$}

\title{NOON-state formation from Fock-state Bose-Einstein condensates}

\affiliation{$^{a}$Centre for Quantum Technologies, National University of Singapore,
Singapore 117543 \\
 $^{b}$ Laboratoire Kastler Brossel, ENS, UPMC, CNRS ; 24 rue
Lhomond, 75005 Paris, France \\
 $^{c}$Department of Physics, University of Massachusetts, Amherst,
Massachusetts 01003 USA}
\begin{abstract}
NOON states (states of the form $|N>_{a}|0>_{b}+|0>_{a}|N>_{b}$ where
$a$ and $b$ are single particle states) have been used for predicting
violations of local realism (Greenberger-Horne-Zeilinger violations)
and are valuable in metrology for precision measurements of phase at
the Heisenberg limit.  We show theoretically how the use of two Fock
state Bose-Einstein condensates as sources in a modified Mach-Zehnder
interferometer can lead to the creation of the NOON state in which $a$
and $b$ refer to arms of the interferometer and $N$ is a subset of the
total number of particles in the two condensates.  The modification of
the interferometer involves making {}``side'' measurements of a few
particles near the sources.  These measurements put the remaining
particles in a superposition of two phase states, which are converted
into NOON states by a beam splitter if the phase states are
orthogonal.  When they are not orthogonal, a {}``feedforward''
correction circuit is shown to convert them into proper form so a NOON
results.  We apply the NOON to the measurement of phase.  Here the
NOON experiment is equivalent to one in which a large molecule passes
through two slits.  The NOON components can be recombined in a final
beam splitter to show interference.
\end{abstract}
\maketitle

\section{Introduction\label{sec:Introduction}}

NOON states are interesting and useful \cite{Dowling08}; they are
{}``all-or-nothing'' states, having the form \begin{equation}
\left|\Phi\right\rangle
=\frac{1}{\sqrt{2}}\left[|N>_{a}|0>_{b}+|0>_{a}|N>_{b}\right]\label{eq:NOONdef}\end{equation} where
the subscripts $a$ and $b$ represent single particle states.
Eq.~(\ref{eq:NOONdef}) represents a superposition of all $N$ particles
in state $a$ and none in $b$, or none in $b$ and all in $a.$ Such
states can have several important applications: 1) Since they are
``Schr\"{o}dinger cat'' states of a system with $N$ particles, one might
use them to demonstrate the quantum interference of macroscopically
distinct objects \cite{Leggett-1}.  2) They can be used to study
violations of quantum realism in the GHZ contradictions
\cite{GHZ,Mermin,Wildfeuer07}.  3) They can be used to violate the
standard quantum limit and attain the Heisenberg limit in metrology by
providing extremely accurate measurements of phase \cite{metrology}.
4) They have been proposed for use in quantum lithography
\cite{Litho}.  NOON states have been made experimentally with $N$ up
to ten particles \cite{Mitchell,Jones,Afek,Walther,Nagata}.  A
two-body NOON state can be constructed by allowing two bosons to
impinge on either side of a 50-50 beam splitter.  This is because the
final state will be a superposition of two-particles on either side of
the splitter according to the Hong\textendash{}Ou\textendash{}Mandel
effect \cite{HOM}.  We have shown previously how one can use two
Bose-Einstein condensate Fock states as sources for an interferometer
that can produce NOON states, with $a$ and $b$ two arms of the
interferometer \cite{CableDowl,QFS}.

The essential idea of the interferometer is as follows. By drawing
off a portion of our condensate particles into a pair of detectors
D1 and D2, we measure phase, which puts the uncounted particles
in a double phase state--a Schr\"{o}dinger cat. If the phase difference
of the two cat branches is $\pi$, then passing the remaining particles
through a beam splitter results in a NOON state in the two output
arms of the interferometer. However, this result occurs only if the
number of particles initially detected is ideal (equal numbers in
D1 and D2). The phase difference can then be adjusted by use of a correction
circuit, in which a ``feedforward'' method sets the transmission
coefficient of a side detector D9 to an appropriate value \cite{CableDowl}.
The result is that a very good approximation to a pure NOON state
can almost always be generated whatever the count of the initial detections
at D1 and D2.

We begin in Sec.~\ref{sec:Interferometer} by introducing the two-stage
interferometer, with several parameters to be determined.  In
Sec.~\ref{sec:Case-with-no}, we look first at the case that D1 and D2
register the same particle number, and show how a NOON state is
obtained at the output by considing the distribution for the relative
phase arising from the initial measurement.  In
Sec.~\ref{sec:The-correction-circuit}, we progress to the general
case, for which D1 and D2  detect different particle numbers.
To obtain a NOON state at the output, a condition is derived relating
the transmission coefficient for D9, and the outcomes at D1, D2 and
D9.  To address the probabilistic nature of the side detections, the
average value for D9 is computed is Sec.~\ref{sec:Finding}, providing
a simple value for the transmission coefficient.  In
Sec.~\ref{sec:The-Corrected-NOON}, the scheme is evaluated by applying
two NOON quality factors to the states at the output.  The
efficiencies of the scheme without and with the correction circuit,
taking into account all possible measurement outcomes, are compared
and explained in Sec.~\ref{sec:Circuit-Efficiency-comparisons}. Finally, applications of the scheme to
phase estimation and to demonstrating quantum interference are given in Sec.~\ref{sec:Applications-of-the}.

\section{Interferometer\label{sec:Interferometer}}

\begin{figure}[h]
\centering \includegraphics[width=4in]{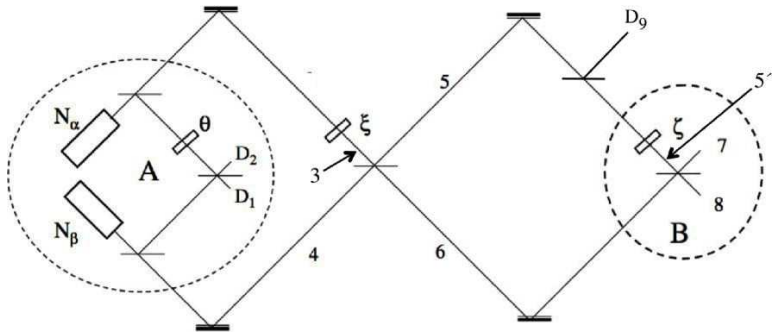}

\caption{An interferometer, in which sources have numbers $N_{\alpha}$ and
$N_{\beta}$, which are partially diverted to interfere at a beam
splitter and detectors D1 and D2. The counts in these detectors are
used to determine the transmission coefficient in the beam splitter
leading to D9 correcting the phases of the beams in arms $5^{\prime}$
and $6$ so that an approximate NOON state emerges 7 and 8. Three
phase shifters $\theta,$ $\xi$, and $\zeta$ must be properly adjusted. }

\label{fig1} 
\end{figure}

The interferometer to be used is shown in Fig.~\ref{fig1}. Two Fock
state sources of number $N_{\alpha}$ and $N_{\beta}$ enter the interferometer.
We will see that the side detectors 1 and 2, situated immediately
after the sources, are a key element; by measuring $m_{1}$ and $m_{2}$
particles in these detectors, the uncounted particles, in arms 3 and
4, are put into phase states. If the phase relation is correct (equal
numbers $m_{1}$ and $m_{2}$ in detectors D1 and D2) then, when these
remaining particles pass through the middle
beam splitter, the result (for suitable value of $\xi$) is a NOON
state in arms 5 and 6. However, if $m_{1}\ne m_{2}$ then we will
show that the transmission coefficient at detector D9 can be adjusted
to a value that corrects the relative phase giving a NOON after the
beam splitter at 7-8.

The annihilation operators at the detectors are found by tracing back
from from a detector to each source. We have

\begin{eqnarray}
a_{1} & = & \frac{1}{2}\left(ie^{i\theta}a_{\alpha}-a_{\beta}\right)\quad\quad\quad a_{2}=\frac{1}{2}\left(-e^{i\theta}a_{\alpha}+ia_{\beta}\right)\nonumber \\
a_{5} & = & \frac{i}{2}\left(ie^{i\xi}a_{\alpha}+a_{\beta}\right)\quad\quad\quad a_{5^{\prime}}=\frac{-\sqrt{T}}{2}e^{i\zeta}\left(ie^{i\xi}a_{\alpha}+a_{\beta}\right)\nonumber \\
a_{6} & = & \frac{1}{2}\left(ie^{i\xi}a_{\alpha}-a_{\beta}\right)\quad\quad\quad a_{7}=\frac{1}{2\sqrt{2}}\left(ue^{i\xi}a_{\alpha}+va_{\beta}\right)\nonumber \\
a_{8} & = & \frac{1}{2\sqrt{2}}\left(ve^{i\xi}a_{\alpha}-ua_{\beta}\right)\quad a_{9}=\frac{-i\sqrt{R}}{2}\left(ie^{i\xi}a_{\alpha}+a_{\beta}\right)\label{eq:Operators}\end{eqnarray}
 where \begin{eqnarray}
u & = & \left(\sqrt{T}e^{i\zeta}-1\right)\nonumber \\
v & = & -i\left(\sqrt{T}e^{i\zeta}+1\right)\end{eqnarray}
 and $T$ and $R=1-T$ are the transmission and reflection coefficients
at the beam splitter leading to D9. We will immediately take $\theta=\pi/2,$
which puts the phase states symmetrically around the zero angle as
we will see.

\section{Case with no correction circuit\label{sec:Case-with-no}}

We first look at the uncorrected situation where we select only
cases in which $m_{1}=m_{2}.$ We then want to show how a NOON state
arises in arms 5 and 6. The amplitude for finding particle numbers
$\{m_{1},m_{2},m_{5},m_{6}\}$ in those detectors is

\begin{equation}
C_{m_{1},m_{2},m_{5},m_{6}}=\left\langle 0\left|\frac{a_{5}^{m_{5}}a_{6}^{m_{6}}a_{1}^{m_{1}}a_{2}^{m_{2}}}{\sqrt{m_{1}!m_{2}!m_{5}!m_{6}!}}\right|N_{\alpha}N_{\beta}\right\rangle \label{C56}\end{equation}
 Put in the forms from Eq.\ (\ref{eq:Operators}) and expand the
binomials to give\begin{eqnarray}
C_{m_{1},m_{2},m_{5},m_{6}} & \sim & \sum_{\{p_{i}\}}\binom{m_{1}}{p_{1}}\binom{m_{2}}{p_{2}}\binom{m_{5}}{p_{5}}\binom{m_{6}}{p_{6}}(-1)^{m_{2}-p_{2}}(ie^{i\xi})^{p_{5}+p_{6}}(-1)^{m_{6}-p_{6}}\nonumber \\
 &  & \times\left\langle 0\left|a_{\alpha}^{p_{1}+p_{2}+p_{5}+p_{6}}a_{\beta}^{m_{1}+m_{2}+m_{5}+m_{6}-p_{1}-p_{2}-p_{5}-p_{6}}\right|N_{\alpha}N_{\beta}\right\rangle \label{eq:Cexpan}\end{eqnarray}
 The second line is evaluated to
 $\sqrt{N_{\alpha}!N_{\beta}!}$$\delta_{p_{1}+p_{2}+p_{5}+p_{6},N_{\alpha}}\delta_{m_{1}+m_{2}+m_{5}+m_{6},N}$,
where $N=N_{\alpha}+N_{\beta}$.
For accurate analysis we could replace one of the summation variables
by use of the $\delta$-function. But for physical analysis we replace
the first by\begin{equation}
\delta_{p_{1}+p_{2}+p_{5}+p_{6},N_{\alpha}}=\int_{-\pi}^{\pi}\frac{d\phi}{2\pi}e^{i(p_{1}+p_{2}+p_{5}+p_{6}-N_{\alpha})\phi}\end{equation}

Putting this into Eq.\ (\ref{eq:Cexpan}) we find that each $a_{\alpha}$
has been replaced by $e^{i\phi}$ and each $a_{\beta}$ by 1. The
result of redoing the sum is then\begin{equation}
C_{m_{1},m_{2},m_{5},m_{6}}=\frac{\sqrt{N_{\alpha}!N_{\beta}!}}{2^{N}\sqrt{m_{1}!m_{2}!m_{5}!m_{6}!}}\int_{-\pi}^{\pi}\frac{d\phi}{2\pi}e^{-iN_{\alpha}\phi}R_{12}(\phi)\left(ie^{i\xi}e^{i\phi}+1\right)^{m_{5}}\left(ie^{i\xi}e^{i\phi}-1\right)^{m_{6}}\label{eq:C56}\end{equation}
 where

\begin{eqnarray}
R_{12}(\phi) & = & (e^{i\phi}+1)^{m_{1}}(e^{i\phi}-1)^{m_{2}}\label{eq:R12}\end{eqnarray}
 Factor out $e^{i\phi/2}$, to give \begin{equation}
R_{12}(\phi)=i{}^{m_{2}}2^{M}e^{iM\phi/2}Q_{12}(\phi)\end{equation}
 where $M=m_{1}+m_{2}$ and\begin{equation}
Q_{12}(\phi)=\left(\cos\frac{\phi}{2}\right)^{m_{1}}\left(\sin\frac{\phi}{2}\right)^{m_{2}}\label{eq:Qphi}\end{equation}

\begin{figure}[h]
 \centering \includegraphics[width=3in]{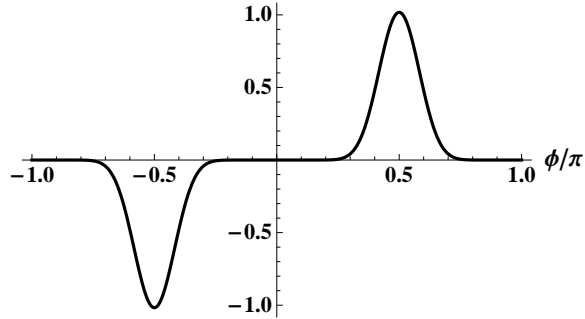} \caption{Plot of $Q_{12}(\phi)$ of Eq.\ (\ref{eq:Qphi}) for
 $m_{1}=m_{2}=15$.
Because we have taken $\theta=\pi/2$ and $m_{1}=m_{2}$ we find the
two peaks at $\pm\pi/2.$ For odd $m_{2}$ we have one positive and
one negative peak. With $m_{2}$ even both peaks are positive. }

\label{Qphi} 
\end{figure}

For arbitrary $m_{1},m_{2}$, the function Q$_{12}$ has peaks at
$\pm\phi_{0}=\pm2\arctan(\sqrt{m_{2}/m_{1}}).$ In the case $m_{1}=m_{2}$
the plot of $Q_{12}$ has peaks at $\pm\pi/2$ as shown in Fig.~\ref{Qphi}.

To test for the presence of the NOON state we approximate the peaks
in $Q_{12}$ of Fig.~\ref{Qphi} by $\delta$-functions:\begin{equation}
Q_{12}(\phi)\sim\delta(\phi-\phi_{0})+(-1)^{m_{2}}\delta(\phi+\phi_{0})\end{equation}
 The result is then\begin{eqnarray}
C_{m_{1},m_{2},m_{5},m_{6}} & \sim & e^{-iN_{\alpha}\phi_{0}}\left(ie^{i\xi}e^{i\phi_{0}}+1\right)^{m_{5}}\left(ie^{i\xi}e^{i\phi_{0}}-1\right)^{m_{6}}\nonumber \\
 &  & +(-1)^{m_{2}}e^{iN_{\alpha}\phi_{0}}\left(ie^{i\xi}e^{-i\phi_{0}}+1\right)^{m_{5}}\left(ie^{i\xi}e^{-i\phi_{0}}-1\right)^{m_{6}}\end{eqnarray}
 When we take $\xi=0$ and $\phi_{0}=\pi/2$ first line is proportional
to $0^{m_{5}}(-2)^{m_{6}}$ requiring $m_{5}=0$ and $m_{6}=N-M$
for non-zero contribution. The second line is proportional to $2^{m_{5}}0^{m_{6}}$
requiring $m_{6}=0.$ Thus we get a NOON state with a superposition
of $\left|m_{1}m_{2}m_{5}m_{6}\right\rangle =\left|M/2,M/2,0,N-M\right\rangle $
and $\left|m_{1}m_{2}m_{5}m_{6}\right\rangle =\left|M/2,M/2,N-M,0\right\rangle $. To make $\phi_{0}=\pi/2$ we must have $m_{1}=m_{2.}$

\begin{figure}[h]
 \centering \includegraphics[width=3in]{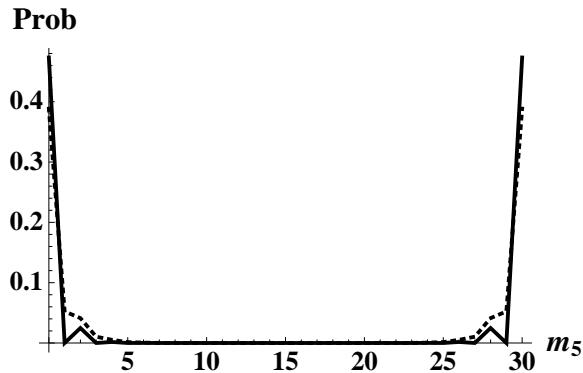}

\caption{Plot of $P_{m_{1},m_{2}m_{5},m_{6}}$ of Eq.\ (\ref{eq:Prob56})
versus $m_{5}$ ($m_{6}=60-m_{5})$ for $N_{\alpha}=N_{\beta}=30,$
$m_{1}=m_{2}=15$, $\xi=0$ (solid line). We also show $m_{1}=18,$
$m_{2}=12$ (dotted line) to illustrate a case where the phase states
are not quite orthogonal. Discrete points are conneced by lines as
a guide to the eye. }

\label{figNOON} 
\end{figure}

To avoid a numerical integral, we use the $\delta$-function generated
in Eq.\ (\ref{eq:Cexpan}) to keep the probability in the form of
a sum:\begin{eqnarray}
P_{m_{1},m_{2}m_{5},m_{6}} & = & Km_{5}!m_{6}!\left|\sum_{p,q,r}\frac{e^{-i(p+q)(\xi+\pi/2)}(-1)^{p+r}}{p!(m_{1}-p)!q!(m_{2}-q)!r!(m_{5}-r)!}\right.\nonumber \\
 &  & \times\left.\frac{1}{(N_{\alpha}-p-q-r)!(p+q+r+m_{6}-N_{\alpha})!}\right|^{2}\label{eq:Prob56}\end{eqnarray}
 By adjusting the phase to $\xi=0$ we get the approximate NOON state as seen
in Fig.~\ref{figNOON}. The NOON state is not perfect because of the
finite width of the peaks in $Q_{12}$. As $m_{1}$  and $m_{2}$
increase the peaks narrow; in the limit in which they can be replaced 
by $\delta$-functions, the output state becomes an ideal NOON. 

Thus we see in this case that we need $m_{1}$ to be only approximately
equal to $m_{2}$ to get a good approximate NOON state.

\section{The correction circuit\label{sec:The-correction-circuit}}

Consider now the complete circuit of Fig.~\ref{fig1}. What do we
have to do to produce a NOON state in D7 and D8 in the case when $m_{1}\ne m_{2}$?
The basic idea to answer this question was given in Ref.~\onlinecite{CableDowl}.
The amplitude for finding the detector counts $\{m_{1},m_{2},m_{7},m_{8},m_{9}\}$
is\begin{eqnarray}
C_{m_{1},m_{2},m_{7},m_{8},m_{9}} & = & \left\langle 0\left|\frac{a_{7}^{m_{7}}a_{8}^{m_{8}}a_{9}^{m_{9}}a_{1}^{m_{1}}a_{2}^{m_{2}}}{\sqrt{m_{1}!m_{2}!m_{7}!m_{8}!m_{9}!}}\right|N_{\alpha}N_{\beta}\right\rangle \nonumber \\
 & \sim & \sum_{\{p_{i}\}}\binom{m_{1}}{p_{1}}\binom{m_{2}}{p_{2}}\binom{m_{7}}{p_{7}}\binom{m_{8}}{p_{8}}\binom{m_{9}}{p_{9}}(-1)^{m_{2}-p_{2}+m_{8}-p_{8}}\nonumber \\
 &  & \times(ie^{i\xi})^{p_{9}}(e^{i\xi})^{p_{7}+p_{8}}u^{m_{8}+p_{7}-p_{8}}v^{m_{7}-p_{7}+p_{8}}\nonumber \\
 &  & \times\delta_{p_{1}+p_{2}+p_{7}+p_{8}+p_{9},N_{\alpha}}\delta_{m_{1}+m_{2}+m_{7}+m_{8}+m_{9},N}\label{eq:C78long}\end{eqnarray}
 Replacing the $\delta$-function by an integral as above gives us
\begin{eqnarray}
C_{m_{1},m_{2},m_{7},m_{8},m_{9}} & = & \frac{e^{i\eta}\sqrt{R^{m_{9}}}\sqrt{N_{\alpha}!N_{\beta}!}}{2^{N}\sqrt{2}^{m_{7}+m_{8}}\sqrt{m_{1}!m_{2}!m_{7}!m_{8}!m_{9}!}}\int_{-\pi}^{\pi}\frac{d\phi}{2\pi}e^{-iN_{\alpha}\phi}R_{129}(\phi)\nonumber \\
 &  & \times\left(ue^{i\xi}e^{i\phi}+v\right)^{m_{7}}\left(ve^{i\xi}e^{i\phi}-u\right)^{m_{8}}\label{eq:C78}\end{eqnarray}
 where $\eta$ is an unimportant phase and\begin{equation}
R_{129}(\phi)=(e^{i\phi}+1)^{m_{1}}(e^{i\phi}-1)^{m_{2}}\left(ie^{i\xi}e^{i\phi}+1\right)^{m_{9}}\label{eq:R129}\end{equation}
 We see immediately that if we take $ie^{i\xi}=1$ then the 9-term
has the same binary form as the 1-term and $m_{1}$ and $m_{9}$ will
simply add. With $\xi=-\pi/2$ we have \begin{equation}
R_{129}=2^{m_{1}+m_{2}+m_{9}}i^{m_{2}}e^{i(m_{1}+m_{2}+m_{9})\frac{\phi}{2}}\left(\cos\frac{\phi}{2}\right)^{m_{1}+m_{9}}\left(\sin\frac{\phi}{2}\right)^{m_{2}}\end{equation}
 which has peaks at $\pm\phi_{0}=\pm2\arctan(\sqrt{m_{2}/(m_{1}+m_{9})}).$

\begin{figure}[h]
 \centering \includegraphics[width=3in]{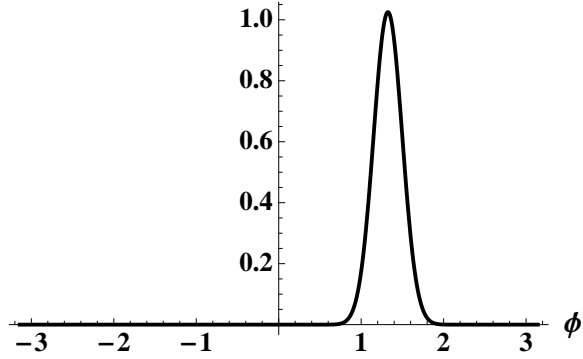}

\caption{Plot of $Q_{129}Q_{8}$ versus $\phi$ showing that it peaks sharply.
The location of the peak is given by the cubic equation given in Eq.
(\ref{eq:Cubic}). Here $m_{1}=22,$ $m_{2}=8$, $m_{9}=14.$ }

\label{fig:Q} 
\end{figure}

However, for arbitrary $\phi$ the second line of Eq.\ (\ref{eq:C78})
becomes (for $e^{i\xi}=-i$), \begin{eqnarray}
\left(-iue^{i\phi}+v\right)^{m_{7}}\left(ive^{i\phi}+u\right)^{m_{8}} & = & (-i)^{m_{7}}\left[\sqrt{T}e^{i\zeta}(e^{i\phi}+1)-(e^{i\phi}-1)\right]^{m_{7}}\left[\sqrt{T}e^{i\zeta}(e^{i\phi}+1)+(e^{i\phi}-1)\right]^{m_{8}}\nonumber \\
 & = & (-i)^{m_{7}}e^{i(m_{7}+m_{8})\phi/2}2^{m_{7}+m_{8}}\left[\sqrt{T}e^{i\zeta}\cos\frac{\phi}{2}-i\sin\frac{\phi}{2}\right]^{m_{7}}\nonumber \\
 &  & \times\left[\sqrt{T}e^{i\zeta}\cos\frac{\phi}{2}+i\sin\frac{\phi}{2}\right]^{m_{8}}\label{eq:L78}\end{eqnarray}
 We can make the factors real if we take $\zeta=\pi/2$, which gives\begin{equation}
C_{m_{1},m_{2},m_{7},m_{8}m_{9}}=\frac{e^{i\eta}\sqrt{R^{m_{9}}}\sqrt{N_{\alpha}!N_{\beta}!}}{\sqrt{2}^{m_{7}+m_{8}}\sqrt{m_{1}!m_{2}!m_{7}!m_{8}!m_{9}!}}\int_{-\pi}^{\pi}\frac{d\phi}{2\pi}e^{i(N_{\beta}-N_{\alpha})\phi/2}Q_{129}(\phi)Q_{8}(\phi)\left[\sqrt{T}\cos\frac{\phi}{2}-\sin\frac{\phi}{2}\right]^{m_{7}}\label{eq:CReal}\end{equation}
 where $\mathcal{M}=m_{1}+m_{2}+m_{9}$, and \begin{equation}
Q_{129}(\phi)=\left(\cos\frac{\phi}{2}\right)^{m_{1}+m_{9}}\left(\sin\frac{\phi}{2}\right)^{m_{2}}\end{equation}
 \begin{equation}
Q_{8}(\phi)=\left[\sqrt{T}\cos\frac{\phi}{2}+\sin\frac{\phi}{2}\right]^{m_{8}}\end{equation}
 $Q_{129}$ peaks sharply at two angles analogous to $Q_{12}$ of
Eq.\ (\ref{eq:Qphi}) and, for large $m_{8}$, $Q_{8}$ peaks sharply
at a different angle; the product also peaks sharply at an intermediate
angle. The product $Q_{129}Q_{8}$ is plotted in Fig.~\ref{fig:Q}
for some specific parameter values to show this. %

If $m_{7}$ is small,  we can assume the last $m_{7}$ factor
in Eq.\ (\ref{eq:L78}) (call it $\Delta(\phi)^{m_{7}})$ is slowly
varying compared to the other factor, which can be approximately represented
as a $\delta$-function at its maximum angle $\phi_{m.}$ Our result
for the probability will then have a factor $\Delta(\phi_{m})^{m_{7}}.$
If we make $\Delta(\phi_{m})$ as small as possible, then the probability
will be small for all $m_{7}$ except for $m_{7}=0$; that will
give us a NOON state.

We can find the maximum angle of $Q_{129}Q_{8}$ by taking the logarithmic
derivative of the quantity. We take $m_{78}$ equal to the total number
of particles entering D7 and D8, that is $m_{78}\equiv m_{7}+m_{8}=N-m_{1}-m_{2}.$
The result of this is the following cubic equation in $X=\tan\frac{\phi_{m}}{2}$:\begin{equation}
(m_{1}+m_{9})X^{3}+\sqrt{T}(m_{1}+m_{9}+m_{78})X^{2}-(m_{2}+m_{78})X-m_{2}\sqrt{T}=0\label{eq:Cubic}\end{equation}
 Also we want $\Delta(\phi_{m})=0$ or \begin{equation}
\sqrt{T}=X\label{eq:Delta}\end{equation}
 Combining the last two equations gives us a value for $T$: \begin{equation}
T=\frac{2m_{2}+m_{78}}{2(m_{1}+m_{9})+m_{78}}\end{equation}
 Since $m_{78}=N-m_{1}-m_{2}-m_{9}$ we have\begin{equation}
T=\frac{N-(m_{1}-m_{2}+m_{9})}{N+(m_{1}-m_{2}+m_{9})}\label{eq:TExact}\end{equation}
 If $m_{1}>m_{2}$ this value of $T$ will be surely less than one.
However, if $m_{2}>m_{1}$ then we will have to go back to Eq.\ (\ref{eq:R129})
and take $\xi=+\pi/2$ to give an appropriate $T$ value, which has
$m_{1}$ and $m_{2}$ interchanged.

Unfortunately, our value of $T$ depends on the value of $m_{9},$
whose probability distribution in turn depends on $T.$ We consider
how to solve this problem in the next section.

\section{Finding $m_{9}$\label{sec:Finding}}

Our expression for $T$ contains $m_{9}$, which we should approximate
in some way to get the best NOON state. We want to set the transmission
coefficient to a value that depends on the count that just occurred
in D1 and D2; of course, then the number going into D9 is probabilistic
and is not precisely known. However we could hope to do well enough
by replacing $m_{9}$ in Eq.\ (\ref{eq:TExact}) by $\left\langle m_{9}\right\rangle $
so \begin{equation}
T=\frac{N-(m_{1}-m_{2}+\left\langle m_{9}\right\rangle )}{N+(m_{1}-m_{2}+\left\langle m_{9}\right\rangle )}\label{eq:Twithm9Ave}\end{equation}
 We then need to find (for fixed  $m_{1}$  and $m_{2}$) $\left\langle m_{9}\right\rangle $,
 which itself depends on $T$.

We proceed just as before to now look at the probability of finding
$m_{5^{\prime}},m_{6},m_{9}$ particles just before the
last beam splitter in Fig.~\ref{fig1}, with  given input values of $m_{1}$  and $m_{2}$. The amplitude for this is\begin{equation}
C_{m_{1},m_{2},m_{5^{\prime}},m_{6},m_{9}}=\left\langle 0\left|\frac{a_{1}^{m_{1}}a_{2}^{m_{2}}a_{5^{\prime}}^{m_{5^{\prime}}}a_{6}^{m_{6}}a_{9}^{m_{9}}}{\sqrt{m_{1}!m_{2}!m_{5^{\prime}}!m_{6}!m_{9}!}}\right|N_{\alpha}N_{\beta}\right\rangle \end{equation}
 We now note that, with the phases chosen, \emph{all} of the operators
are of the form $(a_{\alpha}\pm a_{\beta})$ and for $m_{1}>m_{2}$
we have\begin{equation}
C_{m_{1},m_{2},m_{5^{\prime}},m_{6},m_{9}}=\frac{\sqrt{T^{m_{5}}R^{m_{9}}}}{2^{N}}\left\langle 0\left|\frac{(a_{\alpha}+a_{\beta})^{m_{1}+m_{5^{\prime}}+m_{9}}(a_{\alpha}-a_{\beta})^{m_{2}+m_{6}}}{\sqrt{m_{1}!m_{2}!m_{5^{\prime}}!m_{6}!m_{9}!}}\right|N_{\alpha}N_{\beta}\right\rangle \label{eq:SymmC}\end{equation}
Using this relation we are able to show the following rigorous relation
for the average of $m_{9}$ over a series of measurements at fixed
 $m_{1}$  and $m_{2}$:\begin{equation}
\left\langle m_{9}\right\rangle =(1-T)\left(N-m_{1}-m_{2}-\left\langle m_{6}\right\rangle \right)\label{eq:<m9>}\end{equation}
 Particle conservation requires\begin{equation}
\left\langle m_{5}\right\rangle +\left\langle m_{6}\right\rangle =N-m_{1}-m_{2}\label{eq:<m5+m6>}\end{equation}
 Further we can prove that an extremely good approximation is given
by\begin{equation}
\left\langle m_{5}\right\rangle \approx\frac{m_{1}}{m_{2}}\left\langle m_{6}\right\rangle \label{eq:<m5ovm6>}\end{equation}
 We derive these equations in Appendix A. The physically revelant
solutions to these equations, valid for $m_{1}\ge m_{2}$ are \begin{eqnarray}
T & = & \frac{m_{2}}{m_{1}}\label{eq:Tresult}\\
\left\langle m_{9}\right\rangle  & = & \frac{m_{1}-m_{2}}{m_{1}+m_{2}}(N-m_{1}-m_{2})\label{eq:m9result}\end{eqnarray}
 We can actually compute a rigorous numerical average for $\left\langle m_{9}\right\rangle $
(Appendix B) to show that this formula is quite accurate.

\section{The Corrected NOON State\label{sec:The-Corrected-NOON}}

Now let us return to the expression for the NOON probability and compute
the distribution with the most probable value of $m_{9}$ now known.
One possible formula uses the $\delta$-function in Eq.\ (\ref{eq:C78long})
to eliminate $p_{1}$ and to find the following result for the probability:\begin{eqnarray}
P_{m_{1},m_{2},m_{7},m_{8},m_{9}} & = & \frac{R^{m_{9}}(m_{1}+m_{9})!^{2}m_{7}!m_{8}!N_{\alpha}!N_{\beta}!}{4^{N}2^{m_{7}+m_{8}}m_{1}!m_{9}!}\left|\sum_{p_{2}\cdots p_{8}}\frac{(-1)^{p_{2}}}{(N_{\alpha}-p_{2}-p_{7}-p_{8})!}\right.\nonumber \\
 &  & \times\left.\frac{(i\sqrt{T}-1)^{m_{8}+p_{7}-p_{8}}(i\sqrt{T}+1)^{m_{7}-p_{7}+p_{8}}}{(m_{1}+m_{9}+p_{2}+p_{7}+p_{8}-N_{\alpha})!p_{2}!(m_{2}-p_{2})!p_{7}!(m_{7}-p_{7})!p_{8}!(m_{8}-p_{8})!}\right|^{2}\label{eq:ExactP}\end{eqnarray}
This formula is very general, but has many sums. Nevertheless we can
use the angular expression
given in Eq.\ (\ref{eq:CReal}), which is also exact and leads to fast
and accurate numerical calculations. We have then the alternative formula\begin{eqnarray}
P_{m_{1},m_{2},m_{7},m_{8},m_{9}} & = & \frac{R^{m_{9}}N_{\alpha}!N_{\beta}!}{2^{m_{7}+m_{8}}m_{1}!m_{2}!m_{7}!m_{8!}}\left|\int_{-\pi}^{\pi}\frac{d\phi}{2\pi}e^{-i(N_{\alpha}-N_{\beta})\phi/2}\left(\cos\frac{\phi}{2}\right)^{m_{1}+m_{9}}\left(\sin\frac{\phi}{2}\right)^{m_{2}}\right.\nonumber \\
 &  & \times\left.\left[\sqrt{T}\cos\frac{\phi}{2}+\sin\frac{\phi}{2}\right]^{m_{8}}\left[\sqrt{T}\cos\frac{\phi}{2}-\sin\frac{\phi}{2}\right]^{m_{7}}\right|^{2}\label{eq:P78ByIntegral}\end{eqnarray}

This probability is symmetric in exchange of $m_{7}$
and $m_{8}$; to see this, after the interchange, simply set $\phi=-\phi$
and the result is the same. We summarize the procedure: For a given
$m_{1}$, $m_{2}$ we determine $T$ and $\left\langle m_{9}\right\rangle $
from Eqs.~(\ref{eq:Tresult}) and (\ref{eq:m9result}). The value
of $m_{9}$ used in the probability (other than in the $T$ and $R$
values themselves) should be at or near the most probable value (rounded
to the nearest integer) since the probability distribution is fairly
narrow (see Appendix B).

\begin{figure}[h]
 \includegraphics[width=3in]{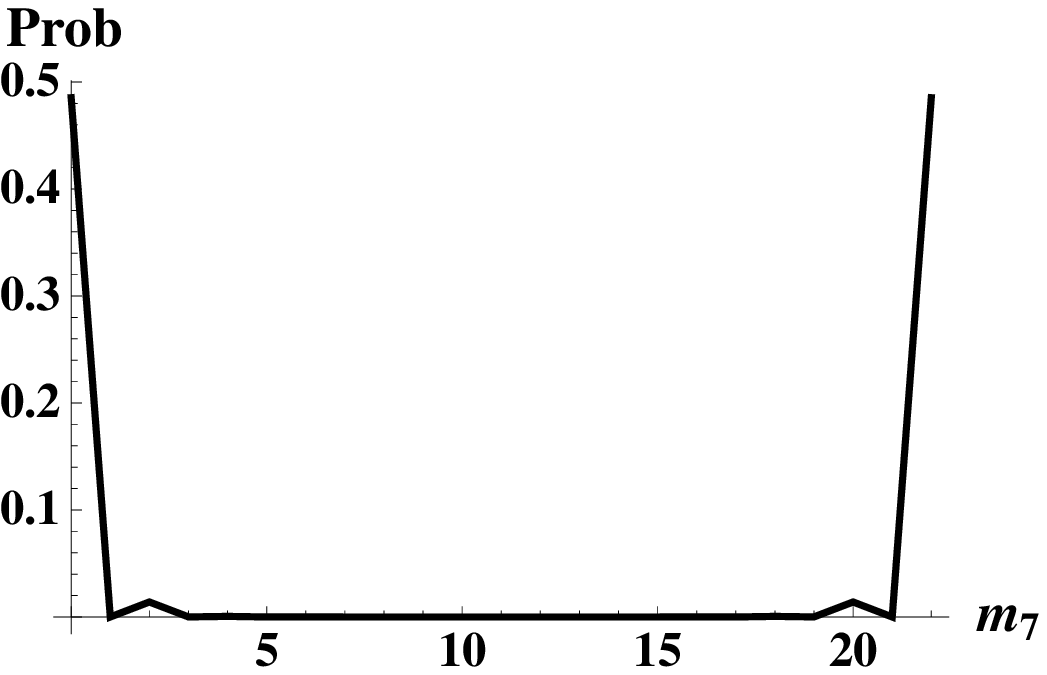}$\quad$\includegraphics[width=3in]{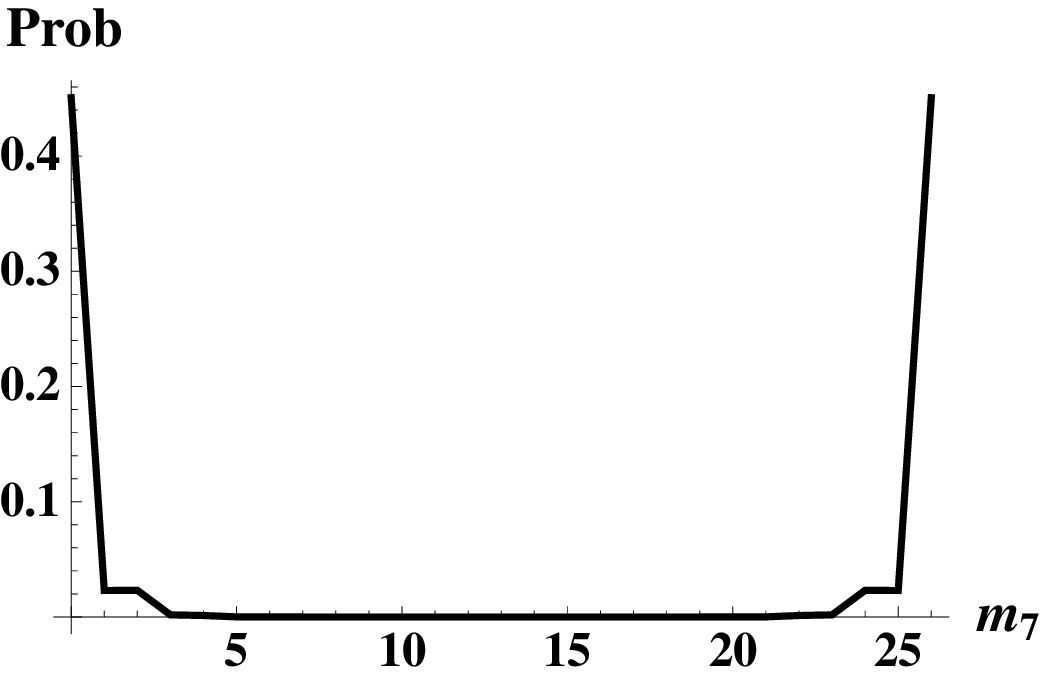}

\caption{$P_{m_{1},m_{2}m_{7},m_{8}m_{9}}$ versus $m_{7}$ when
$\xi=-\pi/2$, $\zeta=\pi/2$ for: (Left) $N_{\alpha}=N_{\beta}=35,$
$m_{1}=22$, $m_{2}=8$ , $m_{9}=18$ (the most probable value),
$T=0.37,$ the value from Eq.\ (\ref{eq:Twithm9Ave}).  Here the NOON
qualities are $q_{1}=0.97$ and $q_{2}=0.99$.  (Right) Same
parameters except with $m_{9}=14.$ We find $q_{1}=0.91$ and
$q_{2}=0.98$.}

\label{fig CablePlots-1} 
\end{figure}

In the first trial we pick the most probable $m_{9}$ value and then
a less probable value of $m_{9}$ with use of the optimum $T$ value.
In the case shown in  Fig.~\ref{fig CablePlots-1} e, $N_{\alpha}=N_{\beta}=35,m_{1}=22,m_{2}=8$,
the value of the transmission is $T=0.36$ and the average value of
$\left\langle m_{9}\right\rangle =18.2$. When $m_{9}=14,$ that is,
a less probable value (the probability of getting this value relative
to the most probable value is $\sim0.4$), the distribution is only
slightly less NOON-like. %

\begin{figure}[h]
 \centering \includegraphics[width=3in]{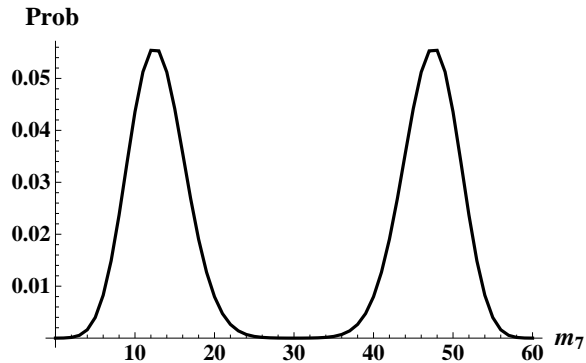}

\caption{Probablity distribution for a poor quality attempted NOON. Here $N=100,$
$m_{1}=35,$ $m_{2}=5$, $T=1$ and $m_{9}=0.$ The NOON quality factors
are $q_{1}=0$ and $q_{2}=0.34.$ }

\label{fig:BadNOON} 
\end{figure}

There is a simple NOON quality factor to test the approximate NOON state,
namely \begin{equation}
q_{1}=2P_{m_{1},m_{2},0,N-m_{1}-m_{2}-m_{9},m_{9}}\end{equation}
that is, twice the value of the probability at $m_{7}=0$; we would
like $q_{1}$ to be as close to 1.0 as possible. The NOON quality 
for the $m_{9}=18$ case in Fig.~\ref{fig CablePlots-1} is 0.97;
for $m_{9}=14$ it is 0.91. Even when the probability of having $m_{7}=0$
or $m_{78}=N-m_{1}-m_{2}-m_{9}$ is high, it is possible to imagine
a peculiar situation where the rest of the particles might be situated
near $m_{78}/2,$ which would diminish the NOON quality of the state.
A quality factor that takes this into account is 
\begin{equation}
q_{2}=4\frac{\Delta^2m_7}{m_{78}^{2}}\label{q2}\end{equation}
where 
 where $\Delta^{2}m_7=\langle m_{7}^{2}\rangle-\langle
m_{7}\rangle^{2}$ is the variance
of $m_{7}$ over the  distribution. In a perfect NOON state the
variance is maximal and 
 $q_{2}=1.$ The worst possible case might be when $P(m_{7}=\frac{m_{78}}{2}-1)=P(m_{7}=\frac{m_{78}}{2}+1)=0.5$
with all others zero; this case has $q_{2}=0.$ A case in which all
probabilities are equal has $q_{2}=0.33.$ The second quality number
seems less sensitive to changes in the resulting NOON states, but
compare them in the case shown in Fig.~\ref{fig:BadNOON}. While
the first quality factor vanishes, the second one is greater than
zero because there is some NOON-like separation in the two peaks.

Next we choose a wide range of values of $m_{1}$, $m_{2}$ to provide
the thorough set of comparisons. The result given in Table \ref{tab:qValues}, which
lists values of the optimal transmission coefficient at $N=140$.
The parameter $m_{78}=m_{7}+m_{8}$ is the total number of particles
involved in the NOON state. In each case the average value of $m_{9}$
is used.

\begin{table} \caption{\label{tab:qValues}Quality factors for a wide range of input $m_{1}$ and $m_{2}$
values, with the resulting transmission coefficients and $m_{9}$ averages
for $N=140$. } \begin{ruledtabular}
    \begin{tabular}{lllllll}
$m_{1}$  & $m_{2}$  & $m_{78}$  & $\left\langle m_{9}\right\rangle $  & $T$  & $q_{1}$  & $q_{2}$\\
\hline
$45$  & 5  & 18  & 72  & 0.11  & 0.976  & 0.990\\
40  & 10  & 36  & 54  & 0.25  & 0.968  & 0.993\\
35  & 15  & 54  & 36  & 0.43  & 0.955  & 0.993\\
30  & 20  & 72  & 19  & 0.67  & 0.932  & 0.992\\
25  & 25  & 90  & 0  & 1.0  & 0.883  & 0.988\\
\end{tabular} \end{ruledtabular} \end{table}

It is likely, in any set of experimental runs, that a random assortment of values of
$m_{1},m_{2}$, and $m_{9}$ will be averaged over in making a NOON
state. What percentage of the inputs will result in good NOON states?
In Appendix C we compare the corrected and uncorrected efficiencies
and show that the correction process is successful in producing good quality NOON states with high
probability.

\section{Applications of the NOON state\label{sec:Applications-of-the}}

\subsection{Metrology}

\begin{figure}[h]
\centering \includegraphics[width=2.5in]{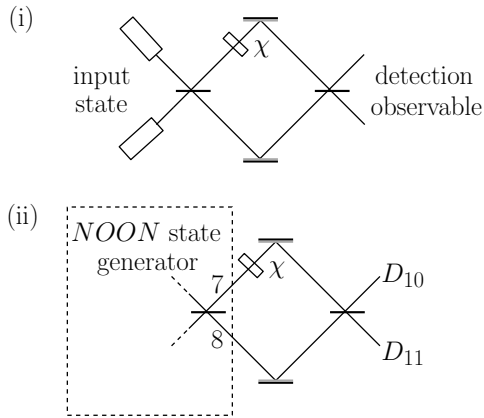} \caption{The application of our NOON-state generator to phase estimation is
illustrated. The aim is determine the unknown path-length difference
in a Mach-Zehnder interferometer with precision approaching the Heisenberg
limit. In (i), we show the usual Mach-Zehnder setup for measurement of
phase.  In (ii), an additional stage is added to the NOON generator
for phase measurements. The NOON is in arms 7 and 8.}
\label{fig:PhaseEstimation} 
\end{figure}

A key application of NOON states is to ultrasensitive sensors, with
fundamental sources of noise reduced to the minimal level permitted by
quantum mechanics \cite{Dowling08,Giovannetti}. Fig.~\ref{fig:PhaseEstimation}(i) 
shows a Mach-Zehnder interferometer set up for the measurement of a path-length difference $\chi
$, a model that can applied to the detection of a variety of physical parameters. A wide
variety of input states, measurement protocols, and decoherence models have
been considered in the literature.  In what follows, we will consider the
usefulness of approximate NOON states generated by the interferometric
method described previously.  The phase estimation process can be
considered as an additional interferometric stage at the output, as
illustrated in  Fig.~\ref{fig:PhaseEstimation}(ii). %

A typical phase estimation procedure is as follows.  For each
estimate of parameter $\chi $, with value $\hat{\chi}$, the experiment is
repeated $t$ times.  It is assumed that the (conditional) probability
distributions $P\left( k|\chi \right) $, for the possible detection outcomes
$k$, are known from theory. Following a Bayesian approach, the posterior
distribution $P\left( \chi |k_{t}\cdots k_{1}\right) $ is obtained from the
prior distribution $P\left( \chi \right) $ by the update rule $P\left( \chi
|k_{t}\cdots k_{1}\right) \propto P\left( k_{t}|\chi \right) \cdots P\left(
k_{1}|\chi \right) P\left( \chi \right) $. A suitable estimator - such as
the mean value for $P\left( \chi |k_{t}\cdots k_{1}\right) $ - is applied to
obtain a value for $\hat{\chi}$. For a so-called ``global'' estimation
procedure no prior information is assumed and $P(\chi )\propto 1/2\pi $; for
a ``local'' estimation procedure, the aim is to track small changes of $\chi $
\cite{Durkin07}. It is for the latter case that NOON states are particularly
useful. A common measure of the statistical information available for a
local estimation procedure is the (classical) Fisher information defined as $%
I_{\text{cl}}=\sum_{k}$ $P\left( k|\chi \right) \left( \frac{d}{d\chi }\ln %
\left[ P\left( k|\chi \right) \right] \right) ^{2}$. Given a total of $\nu
$ independent estimates $\left( \hat{\chi}_{1}\text{,}\cdots \text{,}\hat{%
\chi}_{\nu }\right) $, the Cramer-Rao bound places a lower bound on
precision $\Delta \chi $ (defined as the root mean-square error of the final
estimate) as $\Delta \chi \geq 1/\sqrt{\nu tI_{\text{cl}}}$. This bound
may be assumed to be tight, provided $\nu $ is not too small, although we
will not attempt to provide a detailed statistical analysis on this point. 

Returning to the case of the phase-estimation procedure in Fig.~\ref{fig:PhaseEstimation}(ii), $I_{%
\text{cl }}$is upper bounded by a value termed the quantum Fisher
information $I_{\text{qu}}$. For the case of the interferometric detection
of phase shifts using particle counting, $I_{\text{qu}}$ can be
straightforwardly derived (see Ref.~\onlinecite{Hofmann09}, a special case of the
analysis in Ref.~\onlinecite{Durkin10}), and for the interferometer in Fig.~\ref{fig:PhaseEstimation}(ii) the value
is $I_{\text{qu}}=4\Delta ^{2}m_{7}$ (where $\Delta ^{2}m_{7}$ denotes the
particle-number variance $\left\langle m_{7}^{2}\right\rangle -\left\langle
m_{7}\right\rangle ^{2}$). It is important to verify whether this bound is
tight across the range of possible values for $\chi $ (many existing schemes
are in fact suboptimal in this respect). To check this, we verify that the
approximate NOON states at 7 and 8 satisfy the path-symmetry condition
identified in Ref.~\onlinecite{Hofmann09}. Applying the analysis to the amplitudes for
positions 7 and 8, after $m_{1}$, $m_{2}$ and $m_{9}$ particles are counted
at detectors $D_{1}$, $D_{2}$, and $D_{9}$,
\begin{equation}C_{m_{1},m_{2},m_{7,}m_{8},m_{9}}\propto \left\langle 0\right\vert \frac{%
a_{1}^{m_{1}}a_{2}^{m_{2}}a_{7}^{m_{7}}a_{8}^{m_{8}}a_{9}^{m_{9}}}{\sqrt{%
m_{1}!m_{2}!m_{7}!m_{8}!m_{9}!}}\left\vert N_{\alpha }N_{\beta
}\right\rangle \end{equation}
the path-symmetry condition requires that, 
\begin{equation}C_{m_{1},m_{2},m_{7},m_{8},m_{9}}=\left(
    C_{m_{1},m_{2},m_{8},m_{7},m_{9}}\right)
^{\ast }e^{i\gamma }\end{equation}
where the phase factor $e^{i\gamma }$ is the same for all
possible values for $m_{7}$ and $m_{8}=N-M-m_{7}$ (where $%
M=m_{1}+m_{2}+m_{9} $ and $N=N_{\alpha }+N_{\beta }$), and indices $7$ and $%
8 $ have been swapped. For the amplitudes $%
C_{m_{1},m_{2},m_{7},m_{8},m_{9}} $, this condition can be verified by
inspecting the explicit forms of the operators $a_{i}$, given by
Eq.~(\ref{eq:Operators}),
under (scalar) complex conjugation. As previously it is assumed that ($%
\theta =\pi /2$,$ie^{i\xi }=\pm 1$ and $\zeta =\pi /2$). Since $a_{1}$,
$a_{2}$, and $a_{9}$ are proportional to $a_{\alpha }\pm a_{\beta }$, complex
conjugation  generates only a fixed phase factor contributing to $\gamma $ in
the path-symmetry condition. In addition we find $\left( a_{8}\right)
^{\ast }=ia_{7}$ and $\left( a_{7}\right) ^{\ast }=ia_{8}$. Hence complex
conjugation swaps the $7$ and $8$ labels, and contributes a fixed factor of $%
\left( m_{7}+m_{8}\right) \pi /2$ to $\gamma $.\

To compare the usefulness for phase estimation of states with
the same total particle number\vspace{1pt} $m_{7}+m_{8}$, we adopt the
quality factor of  Eq.~(\ref{q2}) $q_{2}=4\left( \Delta ^{2}m_{7}\right)
/(m_{7}+m_{8})^{2}$; 
$q_{2}$ is normalized between $0$ and $1$, with the maximum value being
attained by a perfect NOON state. For the input state $\left\vert
N_{\alpha }N_{\beta }\right\rangle $ combined at a beam splitter, with $%
N_{\alpha }=N_{\beta }=\left( m_{7}+m_{8}\right) /2$ (assumed to be even), $%
q_{2}$ has the value $1/2+1/(m_{7}+m_{8})$. In particular, there is
already a Heisenberg-limit type scaling for the precision, which at larger
particle numbers is less than that for a corresponding NOON state by a
constant factor of $\sqrt{2}$.

While the direct dual-Fock-state method has a smaller precision by
$\sqrt{2}$ than the corrected NOON method, one must also consider
how many particles the corrected device ``loses'' in the detectors
D1, D2, and D9, especially if the total number of quantum sources
$N$ is limited. If the final output contains a fraction $f$ of the
original source number, $m_{78}=fN$, then to exceed the dual-Fock-state method 
in phase accuracy we require \begin{equation}
\frac{1}{f\sqrt{q_{2}}}\leq \sqrt{2}\end{equation}
For $q_{2}=0.95$ we have $f\geq0.72$. For, say, $N=60$ that would
require $m_{78}\geq44$. We have repeated the averaging calculation
shown in Fig. 9 for $m_{78}=44$ and find in that case $q_{1}=0.81$,
$q_{2}=0.96$ (still a very high value!). However the absolute probability
of that particular value of $m_{78}$ is only $6\times10^{-6}$ compared
to the value $0.02$ for $m_{78}=20$ of Fig. 9. Higher values of
$m_{78}$ become even less likely. Thus the direct dual-Fock-state
process may well provide a more efficient use of limited source resources.

\subsection{Probing the state}

\begin{figure}[h]
 \centering \includegraphics[width=3in]{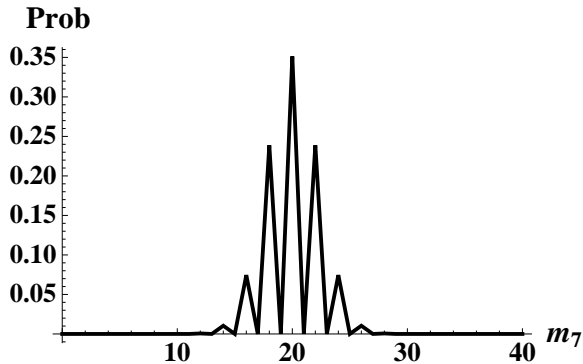}

\caption{Plot of $P_{m_{1},m_{2}m_{7},m_{8}}$ of Eq.~(\ref{eq:P78uzero})
versus $m_{7}$ for $m_{1}=m_{2}=40$, $N_{\alpha}=N_{\beta}=40.$
Here $\zeta=0$ The solid line is the exact result; the dotted line
is the approximation of Eq.~(\ref{eq:P78uzero}). }

\label{figPopOsc} 
\end{figure}

Here we consider the uncorrected case of Sec.~\ref{sec:Case-with-no}
where we took $m_{1}=m_{2}$ so the NOON is formed in arms 5 and 6
and interferes at the last beam splitter with detection in 7 and 8.
We take $\zeta=0$ we also have $\theta=\pi/2$ and $\xi=0$ (as above
in Sec.~\ref{sec:Case-with-no}), in which case we find\begin{equation}
P_{m_{1},m_{2},m_{7},m_{8}}=\frac{K}{m_{7}!m_{8}!}\left|\sum_{p=0}^{m_{1}}\frac{(-1)^{p}}{p!(m_{1}-p)!(N_{\alpha}-p-m_{8})!(m_{2}+m_{8}-N_{\alpha}+p)!}\right|^{2}\label{eq:P78uzero}\end{equation}
where $K$ is a normalization factor. A plot of this probability versus
$m_{7}$ is shown in Fig.~\ref{figPopOsc}. The oscillations are
equivalent to interference fringes and are similar to those found
in Ref.~\onlinecite{MF PRL} where the phase states shown in Fig.~\ref{eq:Qphi}
were allowed to interfere. %

The NOON superposition in the two arms is analogous to macroscopic
particle interference in a two-slit experiment. As explained in more
detail Ref.~\onlinecite{QFS} any attempt to detect the arm in which the
$N$ particles travel results in the destruction of the interference
pattern.

\section{Conclusion}

In conclusion, we have discussed a measurement-based approach to generating
atomic NOON states with a high particle number, developing ideas first
proposed in Refs.~\onlinecite{CableDowl,QFS}. The key requirements
for the method are dual-Fock state Bose-Einstein condensates for the
input, atom interferometry, and particle counting, the basic experimental
feasibility of which have already been demonstrated \cite{BoseInterf,Greiner}.
In the photonic case both high-efficiency number-resolved detection
\cite{Rosenberg} and feed-forward switching \cite{Prevedel} have
been demonstrated experimentally. However photon loss would make successful implementation
of the scheme in a truly scalable manner very difficult \cite{Chen}.
We have shown using two different NOON quality factors that, when
the number of detected particles is sufficiently high, the NOON states
at the output are a very good approximation to the ideal. While dual-Fock states
assumed for the input  enable measurement precision
better than the standard-quantum limit, and Heisenberg-limit-like
scaling, NOON states saturate the Heisenberg limit, and represent
the optimum strategy using nonclassical resources. 
 One can also envisage macroscopic NOON states being
used to demonstrate the quintessential two-slit experiment, in analogy
to a large molecule propagating in a superposition state through a
double slit, and re-interfering with itself at a screen.

In terms of the detailed analysis, the methodology differs somewhat
from that used in Ref.~\onlinecite{CableDowl}. In contrast to the previous work,
an integral representation of $\delta$-functions is used (rather than working in an over-complete
basis of coherent states). This approach is simpler; for example,
the interferometer parameters can be determined by inspection of the
binary form of the annihilation operators at various positions. We
derive rigorously an algebraic condition for the transmission coefficient
for the side-detections for the correction stage. The optimum value
for the coefficient cannot be attained in practice, since it depends
on the unknown outcome at the side detector itself. This problem is
solved by substituting the mean particle count - the best strategy
possible. A simple expression for the transmission coefficient is
derived by considering the mean intensities in different arms of the
interferometer. This leads to a new explanation of the feed-forward
method, and the achievability of high values for the NOON quality factors.

\section*{Acknowledgements} H. C. acknowledges support for this work
by the National Research Foundation and Ministry of Education,
Singapore. \ Laboratoire Kastler Brossel
is \textquotedblleft UMR 8552 du CNRS, de l'ENS, et de
l'Universit\'{e}
Pierre et Marie Curie\textquotedblright .

\section*{Appendix A}

We present here the derivation of Eqs.~(\ref{eq:Tresult}) and (\ref{eq:m9result}).
We start with Eq.~(\ref{eq:SymmC}) and write \begin{equation}
G_{m_{a},m_{b}}=\frac{1}{4^{N}m_{1}!m_{2}!}\left|\left\langle 0\left|(a_{\alpha}+a_{\beta})^{m_{1}+m_{a}}(a_{\alpha}-a_{\beta})^{m_{2}+m_{b}}\right|N_{\alpha}N_{\beta}\right\rangle \right|^{2}\end{equation}
 Then for fixed $m_{5^{\prime}}+m_{6}+m_{9}=N-m_{1}-m_{2}$ we have
\begin{equation}
\left\langle m_{9}\right\rangle =\mathcal{N}\sum_{m_{5^{\prime}},m_{6},m_{9}}\left[\delta_{m_{5^{\prime}}+m_{6}+m_{9},M}m_{9}\frac{T^{m_{5}^{\prime}}R^{m_{9}}}{m_{5^{\prime}}!m_{6}!m_{9}!}G_{m_{5^{\prime}}+m_{9},+m_{6}}\right]\end{equation}
 where $\mathcal{N}$ is a normalization constant. Change summation
variables from $m_{5^{\prime}}$ to $p=m_{5^{\prime}}+m_{9}$. Then
we can pull the $G_{p,m_{6}}$ factor out of the sum on $m_{9}$ to
get\begin{equation}
\left\langle m_{9}\right\rangle =\mathcal{N}\sum_{p,m_{6}}\left\{ \frac{\delta_{p+m_{6},M}}{m_{6}!p!}G_{p+m_{6}}\sum_{m_{9}}\left[m_{9}\frac{p!R^{m_{9}}T^{p-m_{9}}}{m_{9}!(p-m_{9})!}\right]\right\} \end{equation}
 Because $T+R=1$ the sum on $m_{9}$ yields simply $pR$ to give\begin{equation}
\left\langle m_{9}\right\rangle =\mathcal{N}R\sum_{m_{6}}\left\{ \frac{1}{m_{6}!(M-m_{6})!}G_{p+m_{6}}(M-m_{6})\right\} \end{equation}
 An exactly analogous calculation leads to \begin{equation}
\left\langle m_{6}\right\rangle =\mathcal{N}\sum_{p,m_{6}}\left\{ \frac{\delta_{p+m_{6},M}}{m_{6}!p!}m_{6}G_{p+m_{6}}\sum_{m_{9}}\left[\frac{p!R^{m_{9}}T^{p-m_{9}}}{m_{9}!(M-m_{9})!}\right]\right\} \end{equation}
 The $m_{9}$ sum is just $(R+T)^{p}=1$ so that \begin{equation}
\left\langle m_{6}\right\rangle =\mathcal{N}\sum_{m_{6}}\left\{ \frac{m_{6}}{m_{6}!(M-m_{6})!}G_{p+m_{6}}\right\} \end{equation}
 The resulting relation is\begin{equation}
\left\langle m_{9}\right\rangle =(1-T)\left(N-m_{1}-m_{2}-\left\langle m_{6}\right\rangle \right)\end{equation}
 which is Eq.~(\ref{eq:<m9>}).

By particle conservation we have\begin{equation}
\left\langle m_{5}+m_{6}\right\rangle =N-m_{1}-m_{2}\end{equation}
 Note the prime is missing from the $m_{5}$ number here $(m_{5}=m{}_{5^{\prime}}+m_{9})$.

We can also prove a relation between $\left\langle m_{5}\right\rangle $
and $\left\langle m_{6}\right\rangle $. From Eq.~(\ref{eq:C56}),
with appropriate values of the phase shifts entered, we have \begin{eqnarray}
P_{m_{1},m_{2},m_{5},m_{6}} & = & \frac{N_{\alpha}!N_{\beta}!}{4^{N}m_{1}!m_{2}!m_{5}!m_{6}!}\int_{-\pi}^{\pi}\frac{d\phi^{\prime}}{2\pi}\int_{-\pi}^{\pi}\frac{d\phi}{2\pi}e^{-iN_{\alpha}(\phi-\phi^{\prime})}\left[(e^{-i\phi^{\prime}}+1)(e^{i\phi}+1)\right]^{m_{1}+m_{5}}\nonumber \\
 &  & \times\left[(e^{-i\phi^{\prime}}-1)(e^{i\phi}-1)\right]^{m_{2}+m_{6}}\end{eqnarray}
 If we change variables to $\Lambda=(\phi-\phi^{\prime})/2$ and $\lambda=(\phi+\phi^{\prime})/2$
we find \begin{eqnarray}
P_{m_{1},m_{2},m_{5},m_{6}} & = & \frac{N_{\alpha}!N_{\beta}!}{m_{1}!m_{2}!m_{5}!m_{6}!2^{N}}\int_{-\pi}^{\pi}\frac{d\lambda}{2\pi}\int_{-\pi}^{\pi}\frac{d\Lambda}{2\pi}\cos\left[\left(N_{\alpha}-N_{\beta}\right)\Lambda\right]\nonumber \\
 &  & \times\left[\cos\Lambda+\cos\lambda\right]^{m_{1}+m_{5}}\left[\cos\Lambda-\cos\lambda\right]^{m_{2}+m_{6}}\end{eqnarray}
 To find $\left\langle m_{5}\right\rangle $ multiply this by $m_{5}$
and sum over $m_{5}$ and $m_{6}$ subject to the restriction that
$m_{5}+m_{6}=N-m_{1}-m_{2}$; the sum is\begin{eqnarray}
S & = & \sum_{m_{5},m_{6}}\frac{m_{5}}{m_{5}!m_{6}!}\left[\cos\Lambda+\cos\lambda\right]^{m_{5}}\left[\cos\Lambda-\cos\lambda\right]^{m_{6}}\nonumber \\
 &  & \frac{\left(\cos\Lambda+\cos\lambda\right)}{(N-m_{1}-m_{2}-1)!}(2\cos\Lambda)^{N-m_{1}-m_{2}-1}\end{eqnarray}
 This results in the average\begin{equation}
\left\langle m_{5}\right\rangle =K_{m_{1}m_{2}}(m_{1}+1)F(m_{1}+1,m_{2})\end{equation}
 where\begin{eqnarray}
F(m_{1},m_{2}) & = & \frac{1}{m_{1}!m_{2}!}\int_{-\pi}^{\pi}\frac{d\lambda}{2\pi}\int_{-\pi}^{\pi}\frac{d\Lambda}{2\pi}\cos\left[\left(N_{\alpha}-N_{\beta}\right)\Lambda\right]\nonumber \\
 &  & \times(\cos\Lambda)^{N-m_{1}-m_{2}}\left[\cos\Lambda+\cos\lambda\right]^{m_{1}}\left[\cos\Lambda-\cos\lambda\right]^{m_{2}}\end{eqnarray}
 and $K_{m_{1}m_{2}}$ contains other factors unimportant for our
purposes. A similar equation is found for $\left\langle m_{6}\right\rangle $:\begin{equation}
\left\langle m_{6}\right\rangle =K_{m_{1}m_{2}}(m_{2}+1)F(m_{1},m_{2}+1)\end{equation}

For large $N-m_{1}-m_{2}$ the $(\cos\Lambda)^{N-m_{1}-m_{2}}$ factor
peaks very sharply at $\Lambda=0$ and can be replaced by $D_{m_{1}m_{2}}\delta(\Lambda)$,
where $D_{m_{1}m_{2}}$ is a factor that can be lumped into $K_{m_{1}m_{2}}.$
The result is that \begin{eqnarray}
F(m_{1},m_{2}) & \cong & \frac{1}{m_{1}!m_{2}!}\int_{-\pi}^{\pi}\frac{d\lambda}{2\pi}\left[1+\cos\lambda\right]^{m_{1}}\left[1-\cos\lambda\right]^{m_{2}}\nonumber \\
 & = & \frac{1}{m_{1}!m_{2}!}\frac{2^{m_{1}+m_{2}+1}\Gamma(m_{1}+\frac{1}{2})\Gamma(m_{2}+\frac{1}{2})}{(m_{1}+m_{2})!}\end{eqnarray}
 yielding the result \begin{equation}
\frac{\left\langle m_{5}\right\rangle }{\left\langle m_{6}\right\rangle }\cong\frac{m_{1}+\frac{1}{2}}{m_{2}+\frac{1}{2}}\approx\frac{m_{1}}{m_{2}}\end{equation}
 which is Eq.~(\ref{eq:<m5ovm6>}).

\section*{Appendix B}

We want to find a rigorous formula for $\left\langle m_{9}\right\rangle .$
Starting from Eq.~(\ref{eq:SymmC}) we expand the operators, take
matrix elements, yielding a $\delta-$function, which wereplace by
an integral:\begin{eqnarray}
C_{m_{1},m_{2},m_{5^{\prime}},m_{6},m_{9}} & = & \frac{\sqrt{T^{m_{5}}R^{m_{9}}N_{\alpha}!N_{\beta}!}}{2^{N}\sqrt{m_{1}!m_{2}!m_{5^{\prime}}!m_{6}!m_{9}!}}\nonumber \\
 &  & \times\int_{-\pi}^{\pi}\frac{d\phi}{2\pi}e^{-iN_{\alpha}\phi}\left(e^{i\phi}+1\right)^{m_{1}+m_{5^{\prime}}+m_{9}}\left(e^{i\phi}-1\right)^{m_{2}+m_{6}}\\
 & = & \frac{\sqrt{T^{m_{5}}R^{m_{9}}N_{\alpha}!N_{\beta}!}}{\sqrt{m_{1}!m_{2}!m_{5^{\prime}}!m_{6}!m_{9}!}}\\
 &  & \times\int_{-\pi}^{\pi}\frac{d\phi}{2\pi}e^{-i(N_{\alpha}-N_{\beta})\phi}\left(\cos\frac{\phi}{2}\right)^{m_{1}+m_{5^{\prime}}+m_{9}}\left(\sin\frac{\phi}{2}\right)^{m_{2}+m_{6}}\end{eqnarray}
 If $N_{\alpha}=N_{\beta}$ the integral can be done analytically.
We sum the result over all $m_{5^{\prime}}$ and $m_{6}$ to give:\begin{eqnarray}
P_{m_{1},m_{2},m_{9}} & = & \frac{\mathcal{N}R^{m_{9}}}{m_{9}!}\sum_{m_{5^{\prime}}=0}^{N-\mathcal{M}}\frac{T^{m_{5}}}{(m_{5^{\prime}})!(N-\mathcal{M}-m_{5^{\prime}})!}\left\{ \left[1+(-1)^{m_{2}+N-\mathcal{M}-m_{5^{\prime}}}\right]\right.\nonumber \\
 &  & \left.\times\Gamma\left(\frac{1+m_{1}+m_{5^{\prime}}+m_{9}}{2}\right)\Gamma\left(\frac{1+m_{2}+N-\mathcal{M}-m_{5^{\prime}}}{2}\right)\right\} ^{2}\label{eq:P9Prob}\end{eqnarray}
 where $\mathcal{N}$ is a normalization factor.

\begin{figure}[h]
 \centering \includegraphics[width=3in]{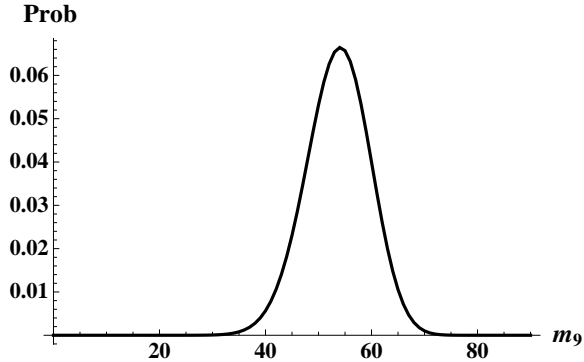}

\caption{$P_{m_{1},m_{2},m_{9}}$ of Eq.~(\ref{eq:Prob56}) versus $m_{9}$
for $N_{\alpha}=N_{\beta}=70,$ $m_{1}=40$, $m_{2}=10$. We find
$\left\langle m_{9}\right\rangle =53.6$, while the approximate formula
of Eq.~(\ref{eq:m9result}) gives 54. }

\label{ExactP9} 
\end{figure}

Fig.~\ref{ExactP9} shows a plot of $P_{m_{1},m_{2},m_{9}}$ of Eq.
(\ref{eq:P9Prob}) for a set of variables having a large value of
$\left\langle m_{9}\right\rangle $. To get this plot we used $T=m_{2}/m_{1}$.
We find the exact result $\left\langle m_{9}\right\rangle =53.6$
compared to the approximation of 54 given by Eq.~(\ref{eq:m9result}).

\section*{Appendix C: Circuit Efficiency comparisons\label{sec:Circuit-Efficiency-comparisons}}

It is likely, in any set of experimental runs, that a random assortment of values of
$m_{1},m_{2}$, and $m_{9}$ will be averaged over in making a NOON
state. What percentage of the inputs will result in good NOON states?
Here we compare the corrected and uncorrected efficiencies to get
an idea of how successful is the correction circuit in providing good
NOON states.

\subsection{Uncorrected Circuit\label{sub:Uncorrected-Circuit}}

\begin{figure}[h]
\centering \includegraphics[width=3in]{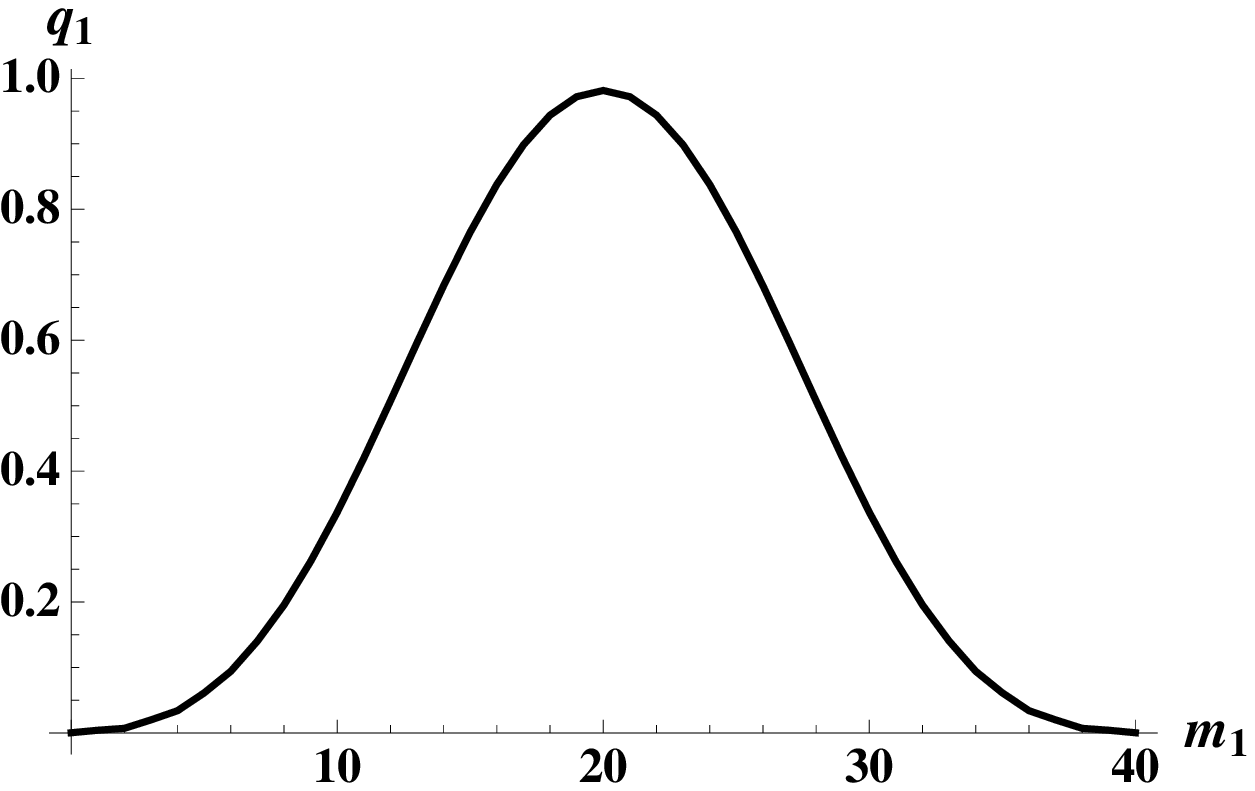}$\quad\quad$\includegraphics[width=3in]{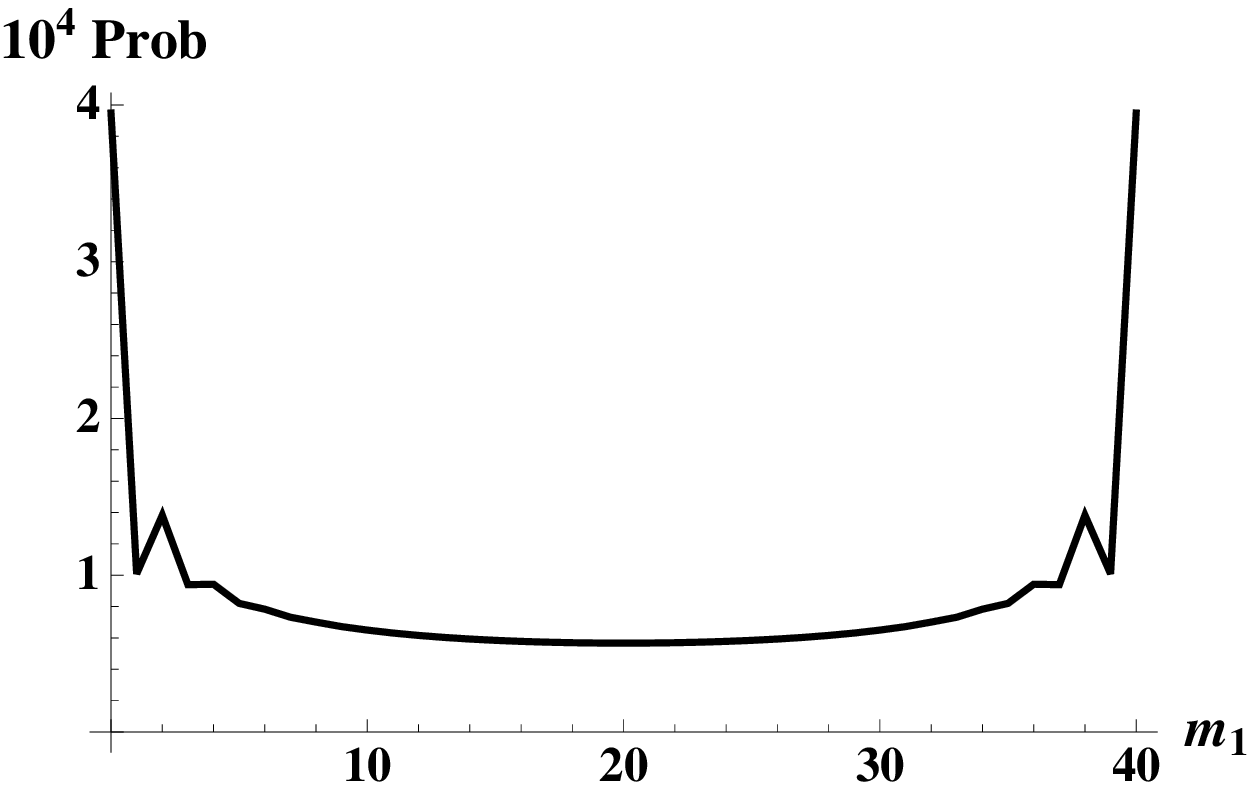}

\caption{Uncorrected circuit quality factor (left) and absolute probability
(right) for fixed $N=60,$ $m_{56}=20$ as a function of $m_{1}$.
The best quality occurs for $m_{1}=m_{2}$. Note that the probability
is smallest for the best quality factor. }

\label{figm56data}
\end{figure}

With the uncorrected circuit one can, for a given $N$ value, compute
in the 2D space of $\{m_{1}, m_{2}\}$, the various possible NOON output
numbers, $m_{56}=N-m_{1}-m_{2}$, and the corresponding NOON quality factor
$q_{1}$ and total absolute probability (probability normalized over all
five $m$ variables) of getting each result. In a 2D plot of the probability
one finds the highest $q_{1}$ factors along $m_{1}=m_{2}$, as expected.
This line of high $q_{1}$ is a minimum of absolute probability --
nonequal $m_{1},m_{2}$ values are more probable. A sample is shown
in Fig.~\ref{figm56data} for $N=60$ and $m_{56}=20.$ Note that
only the middle seven $m_{1}$ values give a quality factor greater
than 0.90 and only the middle three are greater than 0.95. When $m_{56}$
is larger, the quality factors drop so that for, say, $m_{56}=30$
only  the case $m_{1}=m_{2}=15$ gives $q_{1}=0.95$ and only
the middle three points have $q_{1}>0.90$. As $m_{56}$
decreases, the distribution of high quality factors broadens greatly,
but  one is then getting fewer NOON particles as output, and the
absolute probability of occurrence of these states is lower. 

\begin{figure}[h]
\centering \includegraphics[width=3in]{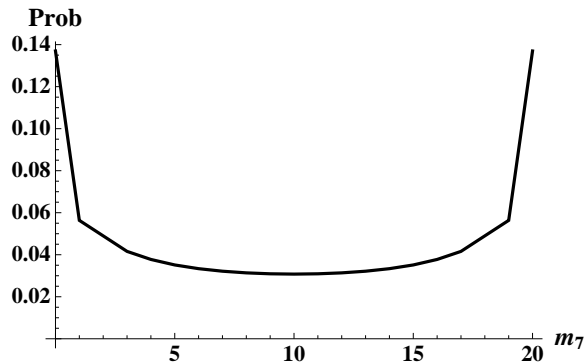}

\caption{The NOON state resulting from averaging over all $m_{1}$ corresponding
to fixed $N=60,$ $m_{56}=20$. The quality factors are low: $q_{1}=0.27$
and $q_{2} = 0.53$. }

\label{fig:NOON56}
\end{figure}

We can average the data of Fig.~\ref{figm56data} over all $m_{1}$ to
get a NOON that would occur if we accepted \emph{all} events at constant
$m_{56}=20.$ The result is shown in Fig.~\ref{fig:NOON56}. The resulting
quality factor is low (0.26) because we have averaged over some very
poor states. The total probability of getting any results corresponding
to $m_{56}=20$ is 0.0036. %

We can assume that someone using the uncorrected circuit would be
more selective and keep only data associated with $m_{1}$ values
that have higher quality factors and sufficiently large $m_{56}$
values. Thus if one demands that the NOON output particle number $m_{56}$
be 20 or larger (still for $N=60$) for $q_{1}=0.90$,  one will get this 6.2\% 
of the
time. Table \ref{tab:MinN} shows the percentages for various output minimum numbers. 

\begin{table} \caption{\label{tab:MinN}Probability $P_{N_{min}}$(in percent) for getting
a NOON with quality $q_{1}$ and output number greater or equal to
$N_{min}$ for $N=60$ for an uncorrected circuit.}
 \begin{ruledtabular}
 \begin{tabular}{ccc}
$N_{min}$& \% ($q_{1}= 0.90$) & \%($q_{1}= 0.95$)\\
\hline
35 & 0.30 & 0\\
30 & 2.7 & 0.2\\ 
20 & 6.2 & 1.6\\
15 & 6.2 & 2.0\\
\end{tabular} \end{ruledtabular} \end{table}

\subsection{Corrected Circuit}

\begin{figure}[h]
\centering \includegraphics[width=3in]{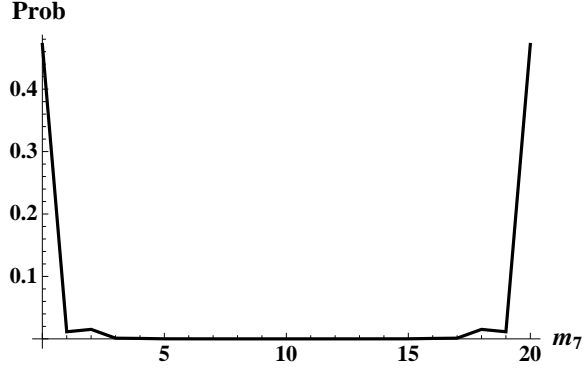}

\caption{The NOON state resulting from averaging over all $m_{1}$ and $m_{2}$
corresponding to fixed $N=60,$ $m_{78}=20$. The $q_{1}$ quality factor is
0.94, while $q_{2}=0.98$ }

\label{fig:NOON78}
\end{figure}

In the corrected circuit we have one more variable $m_{9}$, making
it very difficult to get the data set equivalent to that which led
to, say,  Table \ref{tab:MinN}. However, we can analyze a particular case corresponding
to fixed values of $N$ and output number $m_{78}$. We again choose
$N=60$ and $m_{78}=20.$ First we present a NOON that is a
probability-weighted average
over all $m_{1},m_{2,}$ and $m_{9}$ corresponding to those values.
Of course, we have $m_{9}=N-m_{78}-m_{1}-m_{2}$ determined for each
set of the variables $m_{1}$ and $m_{2.}$ Thus we have a double
sum with the probability distribution given by \begin{equation}
P_{78}=\sum_{m_{1}=0}^{M/2}\left\{
P_{m_{1},m_{1},m_{7},m_{8},m_{9}}+2\sum_{m_{2}=m_{1+1}}^{M-m_{1}}P_{m_{1},m_{2},m_{7},m_{8},m_{9}}\right\} \end{equation}
where $M=N-m_{78}$ and the factor of 2 in the second term takes account
of the symmetry that occurs when $m_{1}$ and $m_{2}$ are interchanged.
The NOON state produced by this process is shown in Fig.~\ref{fig:NOON78}.
The efficiency in getting this high quality result ($q_{1}=0.94$)
is good; that is, output corresponding to any element with this 
output number $m_{78}$ occurs with absolute probability 2.1\%. This NOON quality
 comes without any selection of specially chosen values of $m_{1}$
and $m_{2}$ as occurred in  Table \ref{tab:qValues}. The point is that the $m_{9}$
probability distribution is small away from the high quality points
because the transmission coefficient has been chosen properly. 

\begin{figure}[h]
\centering \includegraphics[width=2.5in]{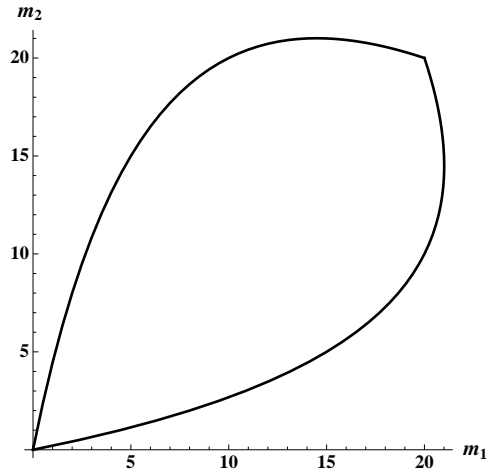}

\caption{Plot of the position in the $\{m_{1},m_{2}\}$ grid at which $m_{9}$
matches its most probable value for $N=60$ and $m_{{78}}=20$. }

\label{fig:m9data}
\end{figure}

\begin{figure}[h]
\centering
\includegraphics[width=3in]{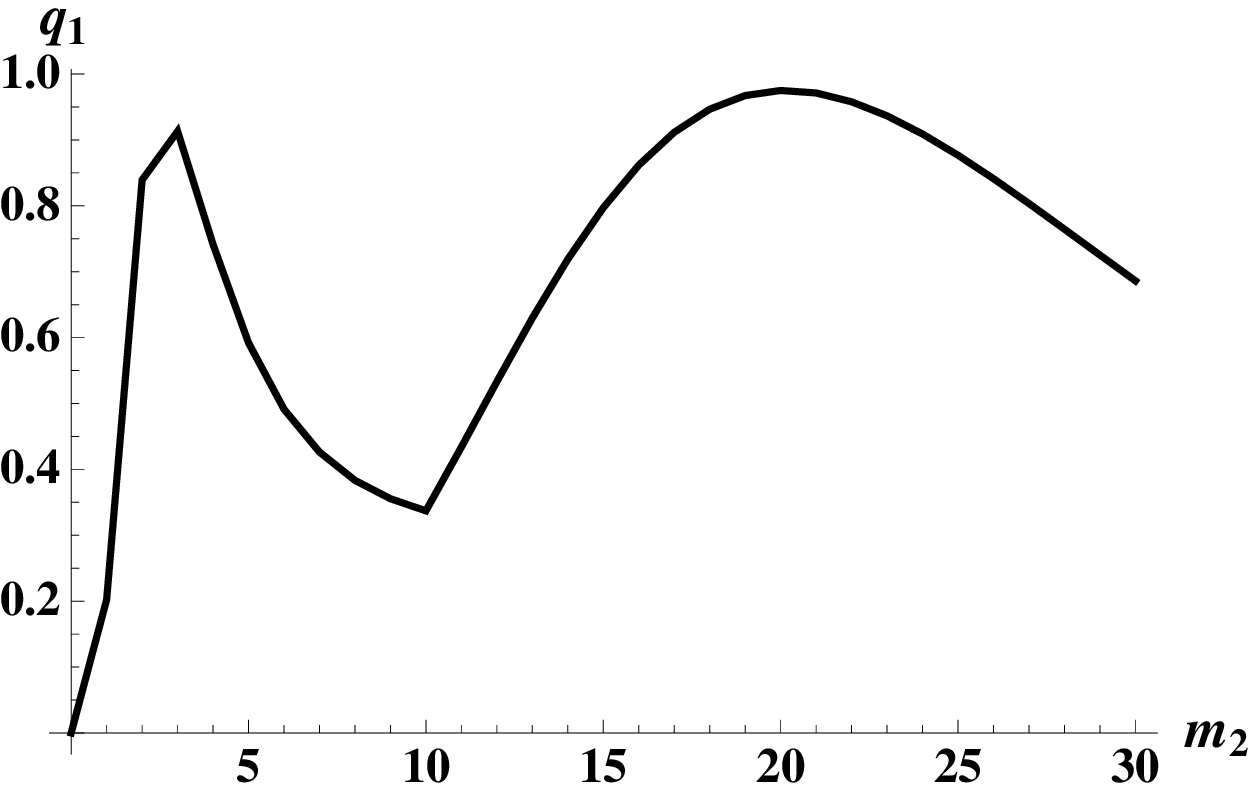}$\quad\quad$\includegraphics[width=3in]{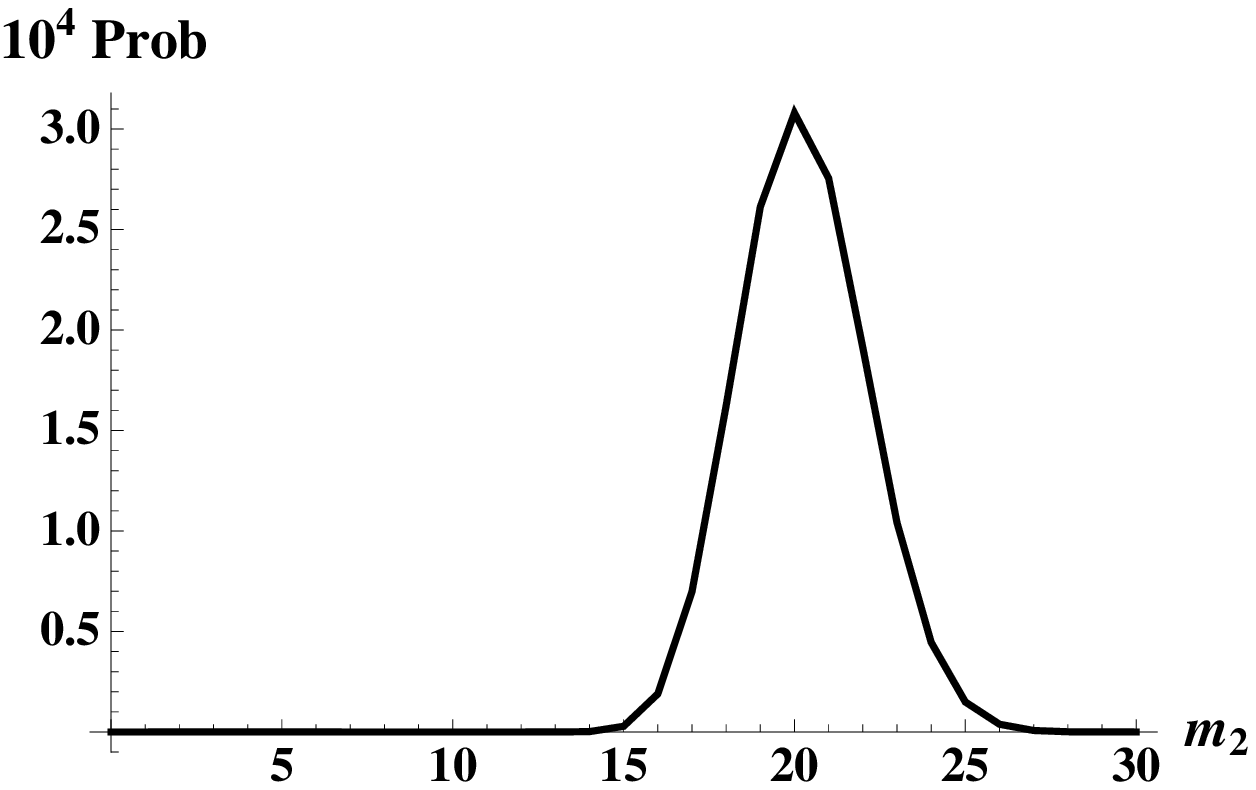}

\caption{Plots of the quality factor $q_{1}$ (left) and absolute probability
(right) versus $m_{2}$ at fixed $m_{1}=10$ for $N=60,$ $m_{78}=20$.}

\label{fig:qAndp10}
\end{figure}

We can see how the probability distribution selects high quality by
taking apart the above NOON output. An important aspect of this is
the position in $\{m_{1},m_{2}\}$ space for which the actual value
of $m_{9}=N-m_{78}-m_{1}-m_{2}$ is equal to its most likely value,
given by Eq.~(\ref{eq:m9result}). This is the position in the space
where the absolute probability will be the largest. Moreover, we expect
that the quality factor will be largest here too. Fig.~\ref{fig:m9data}
shows the plot of the points where this match of the real $m_{9}$
agrees with its most probable value.%
Suppose we now pick a value of $m_{1}$ and plot the quality factor
and probablity versus $m_{2}$: Fig.~\ref{fig:qAndp10} shows these
for $m_{1}=10.$ There are two peaks in the quality factor corresponding
to where $m_{2}$ crosses the places of maximum probability seen in
Figure \ref{fig:m9data}. The cusp occurs at the point $m_{1}=m_{2}=10$
where one switches from having $T=m_{2}/m_{1}$ to the inverse. The
probability shows only the peak at the second crossing; there is also
a peak at the first crossing, but it is too small to appear on the
graph. %
Other samples of such plots show the probability similarly peaking
at the point of largest quality factor. 

\begin{figure}[h]
\centering \includegraphics[width=3in]{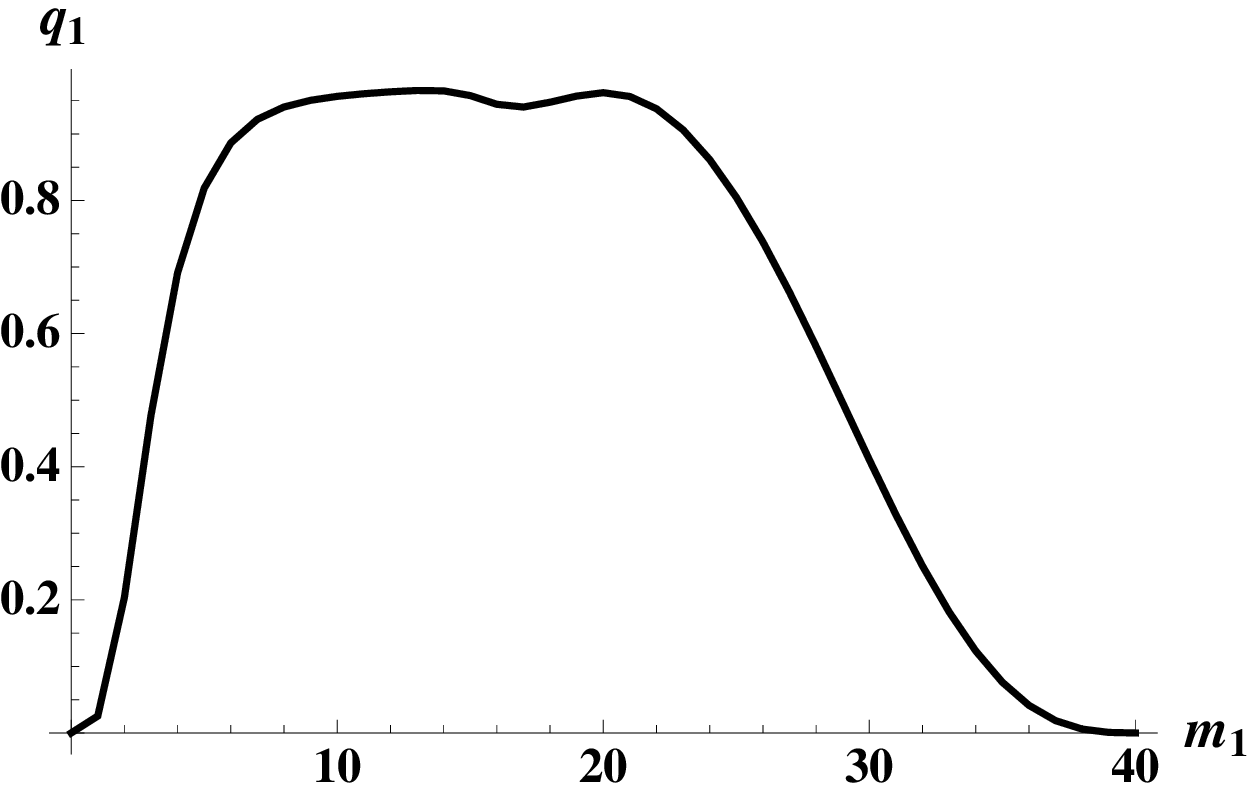}$\quad\quad$\includegraphics[width=3in]{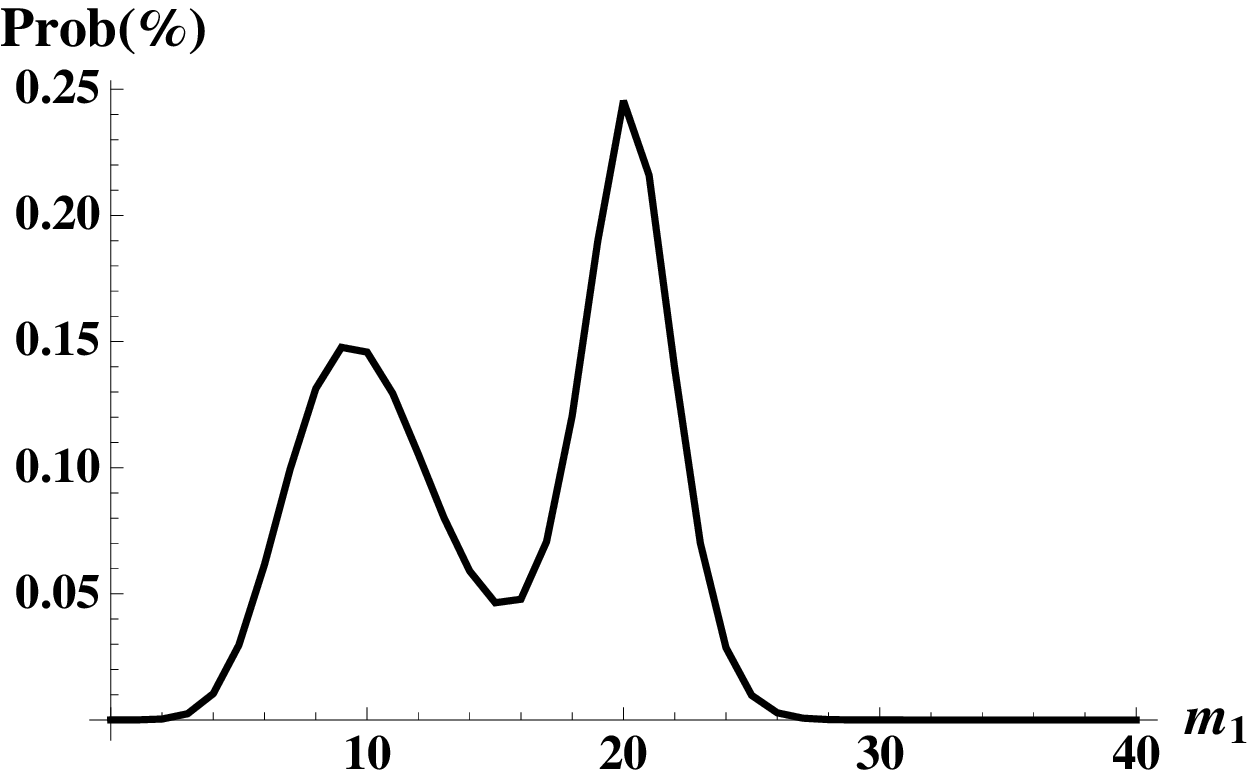}

\caption{Plots of the quality factor $q_{1}$ (left) and absolute probability
(right), summed over all $m_{2}$, versus $m_{1}$ at fixed $m_{1}=20$
for $N=60,$ $m_{78}=20$.}

\label{fig:qAndptotal}
\end{figure}

Finally we plot in Fig.~\ref{fig:qAndptotal} the quality factor and
probablity, \emph{summed} over all $m_{2}$, as a function of $m_{1}$,
again for fixed $N=60,$ $m_{78}=20$. There are two peaks in the
probability since, according to Fig.~\ref{fig:m9data}, there are
equal probability regions in the $\{m_{1},m_{2,}\}$ space symmetrically
at $\{10,20\}$ and $\{20,10\}.$ %

Clearly the correction process is successful in producing good quality
NOON states with high probability.

\end{document}